\documentclass{elsarticle}
\def\comment#1{}

\usepackage{amsmath}
\usepackage{amsfonts}
\usepackage{amssymb}
\usepackage{verbatim}

\usepackage{graphicx}

\usepackage{ulem}
\usepackage{cancel}

\newcommand{\n}{n}
\newcommand{\p}{p}
\newcommand{\rh}{\rho}

\begin{document}

\title{\sf Collective electronic pulsation around giant nuclei in the Thomas-Fermi model}

\author[pes,rom,nic]{H.~Ludwig}
\ead{hendrik.ludwig@icranet.org}

\author[pes,rom,nic]{R.~Ruffini}
\ead{ruffini@icra.it}

\author[pes,rom]{S.-S.~Xue}
\ead{xue@icra.it}

\address[pes]{ICRANet, P.zza della Repubblica 10, I-65122 Pescara, Italy}

\address[rom]{Dipartimento di Fisica and ICRA, Sapienza Universit\`a di Roma
P.le Aldo Moro 5, I-00185 Rome, Italy}

\address[nic]{ICRANet, University of Nice-Sophia Antipolis, 28 Av. de
Valrose, 
06103 Nice Cedex 2, France}

\date{\today}

\vspace{3.0ex}

\begin{abstract}
Based on the Thomas-Fermi solution for compressed electron gas around a giant nucleus, we study electric pulsations of electron number-density,
pressure and electric fields, which could be caused by an external perturbations acting on the nucleus or the electrons themselves. We numerically
obtain the eigen-frequencies and eigen-functions for stationary pulsation modes that fulfill the boundary-value
problem established by electron-number and energy-momentum conservation, equation of state, and
Maxwell's equations, as well as physical boundary conditions, and assume the nucleons in $\beta$-equilibrium at
nuclear density. We particularly study the configuration of ultra-relativistic electrons with a large fraction contained within the nucleus.
Such configurations can be realized for a giant nucleus or high external compression on the electrons.
The lowest modes turn out to be heavily influenced by the relativistic
plasma frequency induced by the positive charge background in the nucleus. Our results can be applied to heavy nuclei in the neutron star crust, as well as to the whole core of a neutron star.
We discuss the possibility to apply our results to dynamic nuclei using the spectral method.
\end{abstract}

%\pacs{97.60.Jd, 21.65.Mn, 03.75.Ss, 26.60.Dd, 26.60.Gj, 26.60.Kp}
%\keywords{Thomas-Fermi model, strong electric field, neutron star dynamics}

\maketitle

\section{\bf Introduction.}\label{intro}

The Thomas-Fermi model that was found independently by Thomas \cite{thomas} and Fermi \cite{fermi} in 1927 quantitatively describes neutral and ionized atoms
of large electron-numbers with great success (see for example Refs.~\cite{pomer1945}-\cite{greiner82}). The Thomas-Fermi solution turns out to be exact when
the electron-number goes to infinity \cite{liebsimon}.
Essentially, the Thomas-Fermi model is a semi-classical and mean-field approach to the problem of many electrons around a nucleus with a large number of
protons. It describes a neutral or charged static equilibrium configuration of electrons around a nucleus with or without compression.
While it turned out to be of limited use in the realm of atomic physics, it has been applied very successfully in astrophysical settings
(see for example Refs.~\cite{ruffinistella80,ruffinistella81,rotondo2,rotondo3,rotondo}).

In this article, on the basis of the Thomas-Fermi solution, namely the equilibrium configurations of electrons compressed around a giant nucleus, 
we investigate radial perturbations (electric pulsations) with spherical symmetry upon the equilibrium configurations.
We find that the spectrum of pulsation modes is determined by two effects: (i) outside the nucleus the speed of sound of the electron gas determines propagation, with possible contributions
from both non- and ultra-realtivistic zones, while (ii) inside the nucleus there is an additional contribution due to the relativistic plasma frequency induced by the nuclear positive charge background.
For sufficiently low frequency modes this leads to the perturbation dying away exponentially within the volume of the nucleus, rendering it effectively unavailable for wave propagation.
To study the configuration of ultra-relativistic electrons with a large
fraction contained within the nucleus, we choose a proton number $Z=10^6$ for the purpose of practical numerical simulation and illustration of (ii).
While the effects we observe are also present at smaller $Z\approx 10^3-10^4$, a more realistic configuration that might be expected in the very deep crust of neutron stars in the form of
\textit{pasta equation of state}, they are less pronounced (see Figure \ref{fig:10to2and4SpectPlot}),
and high electron densities partially rely on the gravitational pressure in this case. Instead we choose $Z=10^6$ because here most electrons are kept inside the nucleus solely by the electric interaction,
and $\beta$-equilibrium is saturated throughout the nucleus. For the effects we observe it is essential that electron densities approach proton densities inside the nucleus, in any other case including
high pressure laboratory setups, the spectrum of the vibrational modes would be dominated by the equation of state and the corresponding speed of sound of the electron gas, the only feature being
a transition from non- to ultra-relativistic conditions (see discussion in the conclusions and Figure \ref{fig:10to2and4SpectPlot}).

The electrons around a static nucleus are treated as a perfect
fluid described by thermodynamic number-density 
$\n$, energy-density $\rh$ and pressure $\p$ with non-vanishing electric potential and field. In addition to the equation of state at zero temperature,
these physical quantities fully obey the Maxwell field-equation, Euler equation and the first thermodynamical law that follows from electron-number and energy-momentum conservations.
This system is completely determined with appropriate physical boundary conditions.
In order to study the perturbative electric pulsations, we have linearized these relations and equations, based on the prescription of Eulerian
and Lagrangian perturbations of the Thomas-Fermi equilibrium configuration. As a result we obtain a homogeneous second-order differential equation for perturbations
satisfying appropriate physical boundary conditions. 

As a first step, we focus on the stationary solution $(\propto e^{i\omega t})$ with the characteristic eigen-frequencies $\omega$ of electric perturbations 
(pulsations) of the Thomas-Fermi system, so as to understand what are time-scales (inverse frequencies) at which the system responds to external actions. In a future work we are planning to make use
of these results to solve the problem of a general perturbation of the nucleus by means of the spectral method.
Numerically and analytically solving this well-defined eigenvalue problem in a hybrid approach, we have obtained the eigen-frequencies and eigen-functions for electric perturbations of the Thomas-Fermi
equilibrium configurations in spherical symmetry. Our study is an analogy of the classical investigation of stellar stability against gravitational
pulsations \cite{chandra,wheeler}, and also in the context of the Thomas-Fermi model for atoms similar calculations have been performed \cite{greiner1,greiner2,wheeler3}. Our work is distinct because we consider a giant
nucleus embedded in electron gas, and pay close
attention to its interior, where the positive charge background of the nucleus induces a plasma frequency for electron oscillation. The purpose of this study is the application to astrophysical systems composed
of nuclear matter and electrons, specifically to dynamical phenomena that could lead to creation of strong electric fields. We have drawn our attention to the ultra-relativistic limit corresponding to the polytropic equations of state
($\p \propto \n^{\Gamma_1}$) with adiabatic index $\Gamma_1=4/3$. Our model could, however, easily be extended to include the transition to non-relativistic regimes, which are reached at compression
radii one order of magnitude higher than the one ($r_\mathrm{max}=1000\,\lambda_\pi$) we are considering here (see Figure \ref{fig:equilibriumPlot}). We present discussions on the physical relevance of our results
and possible experimental testability in the lower crust and core of neutron stars.

The article is organized as follows: In Sec.~\ref{basics}, we present fundamental equations, including the equation of state, describing the dynamics of an electron fluid in electromagnetic fields. 
In Sec.~\ref{thomFer}, we show that in the static case these equations are reduced to the Thomas-Fermi equation, and present its solution. In Secs.~\ref{pert} and \ref{exSol},
we discuss the linear perturbations of the Thomas-Fermi solution, and derive the dynamical equations and boundary conditions that govern these perturbations, as well as obtain their
stationary solutions by numerical and analytical methods. In the final Sec.~\ref{conClu},  we present some discussions on the physical relevance of our results and the possible
observational effects. We use the electron charge $-e$, natural units with $\hbar = c = 1$ and Gauss convention for electrodynamics throughout the article, unless otherwise specified.

\section{\bf Basic Equations of a perfect electron fluid}\label{basics}

In the Thomas-Fermi model electrons around a nucleus are described as a perfect electron-fluid with the energy-momentum tensor (see for example Ref.~\cite{Weinberg1972})
\begin{eqnarray}
T^{\mu\nu} &=& \p g^{\mu\nu}+( \p+\rh)U^\mu U^\nu,
\label{etensor}
\end{eqnarray}
and the chemical potential $\mu=(\rh+\p)/\n$, where $\rh$, $\n$ and $\p$ are the energy, and number densities and pressure of the electron-fluid
in the comoving frame. The motion of a fluid-element is described by its four velocity $U_\mu$. The basic equations that govern the system are
the electron-number conservation (electric current $J^\mu=-e \n U^\mu$ continuity equation): 
\begin{align}\label{enNumCons}
 \nabla_\mu (\n U^\mu)= 0=\nabla_\mu J^\mu, \end{align}
and energy-momentum conservation: (i) along the fluid flow line (see for example Refs.~\cite{wilson,hrx2012}),
\begin{align}\label{eafEqn}
U^\mu\nabla_\nu T^{\nu}_{\,\,\,\mu} = -U^\mu F_{\mu\nu}J^{\nu}=0,
\end{align}
(ii) in the orthogonal direction to the flow line (see for example Refs.~\cite{wheeler2,brown}),
\begin{align}\label{eulEqn}
(g_{\nu\lambda}+U_\nu U_\lambda)\nabla_\mu T^{\mu\lambda}=\mu\,\n\,U^\mu \nabla_\mu U_\nu+(g_\nu{}^\mu+U_\nu U^\mu)\partial_\mu \p=e\,F_{\mu\nu}\n\,U^\mu,
\end{align}
which is the Euler equation of the electron-fluid. In addition, the Maxwell field equations can be summarized by the divergence equation
\begin{align}\label{maxwell}
 \nabla_\mu F^{\mu\nu}&=4\pi J^\nu=-4\pi e \n U^\nu
\end{align}
and the Bianchi identity $\nabla_{[\mu}F_{\nu\lambda]}=0$, where the electromagnetic field tensor is $F^{\mu\nu}=\nabla_\mu A_\nu-\nabla_\nu A_\mu$.  Using the covariant derivative $\nabla_\mu$, we 
work in non-Cartesian, spherical coordinates to explicitly express components of these equations.

Equation (\ref{eafEqn}) is equivalent to the first law of thermodynamics along the flow line of the perfect electron. Eqs. (\ref{enNumCons}) and (\ref{eafEqn}) together
imply (see for example Refs.~ \cite{Weinberg1972,wilson})
\begin{align}
 \frac{d\rh}{d\tau}=\frac{\p+\rh}\n\frac{d \n}{d\tau}.
\end{align}
The pressure, number- and energy-densities of the perfect electron-fluid are given by 
\begin{align}
\n&=2\int_0^{P^F(r)}\frac{\mathrm{d}^3 {\bf p}}{(2\pi)^3}=\frac{m_e{}^{\!3}\eta^3}{3\pi^2},\label{fermiNumDens}\\ \nonumber\\
\p&=\frac{2}{3}\int_0^{P^F(r)}\frac{\mathrm{d}^3 {\bf p}}{(2\pi)^3}\left(\frac{{\bf p}^2}{p_0}\right)
=\frac{m_e{}^{\!4}}{24\pi^2}\left[\left(2 \eta^3-3 \eta\right)\sqrt{\eta^2+1}+3\,\mathrm{arsinh}{\,\eta}\right],\label{fermiPress}
\\ \nonumber\\
\rh&=2\int_0^{P^F(r)}\frac{\mathrm{d}^3 {\bf p}}{(2\pi)^3}\,p_0
 =\frac{m_e{}^{\!4}}{8\pi^2}\left[\left(2 \eta^3+\eta\right)\sqrt{\eta^2+1}-\mathrm{arsinh}{\,\eta}\right],\label{fermiEn}
\end{align}
with the Fermi momentum $P_F=(3 \pi^2 \n)^{1/3}$ and its dimensionless form $\eta=P_F/m_e$, the energy spectrum of a free electron $p_0=({\bf p}^2+m_e^2)^{1/2}$ as well as electron momentum ${\bf p}$ and mass  $m_e$.
In the non-relativistic limit $\eta\rightarrow 0$, the electron energy density and pressure have the expansions
\begin{align}
% \rh\approx \frac{m_e (P_F)^3}{3 \pi^2}+\frac{(P_F)^5}{10 \pi^2 m_e}\ \ \ \ \ \ \mathrm{and}\ \ \ \ \ \ \p\approx \frac{(P_F)^5}{15 \pi^2 m_e}.\label{nonre}
\rh\approx \frac{m_e{}^{\!4}}{\pi^2}\left(\frac{\eta^3}{3}+\frac{\eta^5}{10}\right)\ \ \ \ \ \ \mathrm{and}\ \ \ \ \ \ \p\approx \frac{m_e{}^{\!4}}{\pi^2}\frac{\eta^5}{15}.\label{nonre}
\end{align}
In the ultra-relativistic limit $1/\eta\rightarrow 0$ the electron energy density and pressure have the expansions
\begin{align}
% \rh\approx \frac{(P_F)^4}{4 \pi^2}\ \ \ \ \ \ \mathrm{and}\ \ \ \ \ \ \p\approx \frac{(P_F)^4}{12 \pi^2}.\label{ultra}
\rh\approx \frac{m_e{}^{\!4}}{\pi^2}\frac{\eta^4}{4}\ \ \ \ \ \ \mathrm{and}\ \ \ \ \ \ \p\approx \frac{m_e{}^{\!4}}{\pi^2}\frac{\eta^4}{12}.\label{ultra}
\end{align}
For such a system of electron gas, the fully relativistic equation of state (EOS) is exactly determined by Eqs.~(\ref{fermiPress}) and (\ref{fermiEn}), and
\begin{align}\label{adiaInd}
% \Gamma_1 \equiv \frac{\partial\ln \p}{\partial\ln \n}=\frac{8(P_F)^5}{3 P_F\left[2(P_F)^2-3 m_e^2\right]\left[m_e^2+(P_F)^2\right]+9 m_e^4 \sqrt{m_e^2+(P_F)^2}\mathrm{arcsch}\ m_e/P_F},
 \Gamma_1 \equiv \frac{\partial\ln \p}{\partial\ln \n}=\frac{8\,\eta^5}{6\,\eta^5-3\,\eta^3-9\,\eta+9\,\sqrt{\eta^2+1}\,\mathrm{arsinh}\,\eta},
\end{align}
stands for the adiabatic index with $\p\propto \n^{\Gamma_1}$.
In the non-relativistic limit, the adiabatic index $\Gamma_1= 5/3$. In the ultra-relativistic limit $\Gamma_1= 4/3$. For detailed discussions of non- and
ultra-relativistic limits, see Refs.~\cite{Weinberg1972,landau}.

\section{\bf The relativistic Thomas-Fermi model}\label{thomFer}

In the following sections, we will treat the electric pulsation phenomenon as a linear perturbation of an equilibrium configuration of electrons surrounding a finite size nucleus.
To obtain the equilibrium configuration, we adopt the Thomas-Fermi model and assume that the nucleus has homogeneous density and charge distributions.
First we present the Thomas-Fermi equation in its full relativistic form, from which we then derive its ultra-relativistic limit. In the spherically symmetric case,
the fully relativistic Thomas-Fermi model at zero temperature is given by \cite{thomas,fermi,fmt1949}
\begin{align}\label{relThomFerAnsatz}
 E^F=m_e \left(\sqrt{\eta^2+1}-1\right)-e V(r),
\end{align}
and the Poisson equation
\begin{align}\label{posequ}
\nabla_\mu F^{\mu 0}=\nabla_a E^a=-\nabla^2 V(r)=4\pi e (n_p-\n),
\end{align}
which is the static limit of Maxwell's equation (\ref{maxwell}) with four potential $A_\mu=(V(r),0,0,0)$ and four velocity $U_\mu=(1,0,0,0)$. The proton density $n_p$ is assumed to be homogenous within the nucleus
and zero outside, and to possess no dynamics.
In addition to the equivalence of Poisson's equation (\ref{posequ}) to Maxwell's equation (\ref{maxwell}) under our adopted symmetries, we want to show the equivalence of Eq.~(\ref{relThomFerAnsatz}) to the static
version $(U_\mu=(1,0,0,0))$
of the Euler equation (\ref{eulEqn}) for the electron fluid at zero temperature. Here the only space-dependent quantities
are the Fermi momentum $P^F(r)$ and the electromagnetic potential $V(r)$. Using $V'(r)=-E(r)$, the first $r$-derivative of Eq.~(\ref{relThomFerAnsatz}) becomes
\begin{align}\label{relThomFerAnsatzDer}
 \frac{m_e \eta\,\eta'}{\sqrt{\eta^2+1}}+e E=0.
\end{align}
This should coincide with the aforementioned static Euler equation, which is the time-independent version of Eq.~(\ref{eulEqn}) and reads
\begin{align}\label{staticEuler}
 \frac{\p'}\n+e E=0.
\end{align}
From Eq.~(\ref{fermiPress}) for the pressure, we calculate
\begin{align}\label{pressureGradientFermi}
 \p'(r)=\frac{1}{3\pi^2}\frac{m_e{}^{\!4}\eta^4\,\eta'}{\sqrt{\eta^2+1}}=\frac{\n\,m_e\,\eta\,\eta'}{\sqrt{\eta^2+1}},
\end{align}
where we used $n=m_e{}^{\!3}\eta^3/3\pi^2$ in the last step. Plugging (\ref{pressureGradientFermi}) into (\ref{staticEuler}) we obtain Eq.~(\ref{relThomFerAnsatzDer}).
In conclusion, we proved that the Thomas-Fermi model is equivalent to the particular static versions of Euler's equation (\ref{eulEqn}) and Maxwell's equation (\ref{maxwell}). 

We will express all quantities in the scale of the reduced pion Compton length $\lambda_\pi=1/m_\pi$ throughout the paper.
Eqs.~(\ref{posequ}) and (\ref{relThomFerAnsatzDer}) become (see Refs.~\cite{rotondo2,rotondo3,rotondo})
\begin{align}\label{fullrelFermi}
 \frac{1}{3 r}\frac{\mathrm{d}^2\chi(r)}{\mathrm{d}r^2}=-\frac{\alpha}{\Delta^3}\theta(r_c-r)+\frac{4\alpha}{9\pi}\left[\frac{\chi^2(r)}{r^2}+2\,m_e\frac{\chi(r)}{r}\right]^{3/2},
\end{align}
where the fine-structure constant $\alpha\equiv e^2/4\pi$,  the function $\chi(r)/r\equiv e V(r)+E^F$, and the nuclear radius $r_c\equiv \Delta\, Z^{1/3}\,$. Within this radius, we assume
a positive charge density of $n_p=3/(4\pi\Delta^3)$.
The number of protons inside the nucleus is denoted by $Z$, and $\Delta$ is a quantity with dimension of length that we use to parametrize
the positive charge density. Assuming nuclear density and $Z\approx A/2$ as is the case for common nuclei, one obtains $\Delta\approx 1$. In our case, however, the electrons occupy basically the same volume
as the nucleons, and assuming nuclear density together with $\beta$-equilibrium leads to $\Delta\approx 2.85$. The Heavyside function $\theta(r)$ is used to model a homogeneously charged nucleus.
Moreover, we consider an external pressure, 
i.e.~all electrons are compressed around the nucleus in a spherical region with a finite radius $r_\mathrm{max}$  in order to obtain a discrete spectrum. To arrive at the uncompressed case, one would have to consider the limit
$r_\mathrm{max}\rightarrow\infty$. Therefore, Eqs.~(\ref{relThomFerAnsatz}) and (\ref{fullrelFermi}) have
to be solved with boundary conditions (see Refs.~\cite{rotondo2,rotondo3,rotondo,fmt1949})
\begin{align}\label{equilBoundCon}
\chi(0)=0\mathrm{\ \ \ and\ \ \ }\chi(r_\mathrm{max})=r_\mathrm{max}\,\chi'(r_\mathrm{max}),
\end{align}
which correspond to a finite Fermi momentum (or number-density) at the origin and a vanishing pressure gradient (\ref{pressureGradientFermi}) at the boundary, respectively.
At the boundary, we impose that the electrostatic potential $V(r_\mathrm{max})$ and field $E(r_\mathrm{max})$ vanish, so the Fermi energy is given by
\begin{align}
E^F=\chi(r_\mathrm{max})/r_\mathrm{max}=\chi'(r_\mathrm{max}).\label{featb}
\end{align} 
The number-density can be recovered via
\begin{align}\label{fullrelNumDens}
 \n=\frac{1}{3\pi^2}\left[\frac{\chi^2}{r^2}+2\,m_e \frac{\chi}{r}\right]^{3/2}.
\end{align}

\begin{figure}
 \centering
\includegraphics[scale=0.7]{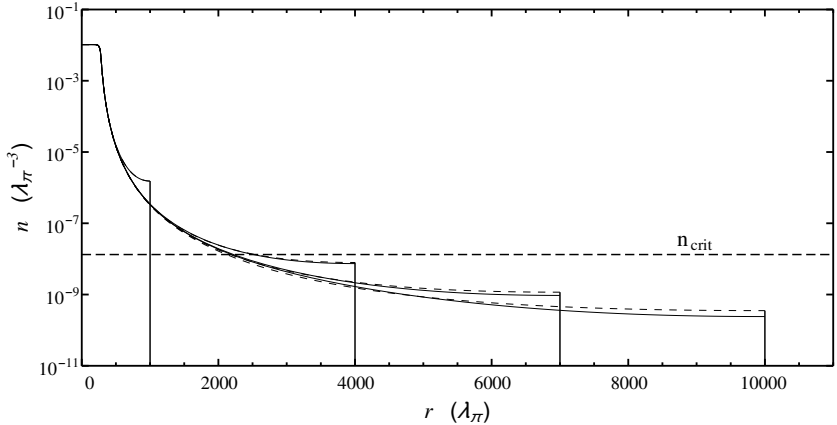}
 \caption{Equilibrium number density for $Z=10^6$, $\Delta=2.85$, $r_\mathrm{nuc}=285$, and compression radii for the electron gas of $r_\mathrm{max}=1000,\,4000,\,7000,\,10000$. The solid lines depict
 the full relativistic solution, and the dashed lines the ultra-relativistic approximation. The horizontal dashed line indicates the transition from ultra- to non-relativistic densities at $n_\mathrm{crit}=(2m_e)^3/(3\pi^2)$.}
 \label{fig:equilibriumPlot}
\end{figure}

\section{\bf The linear perturbation}\label{pert}

We turn to study the main subject of this article: the electric pulsation of a compressed electron gas around a giant nucleus.
In spherical symmetry, this perturbation is described by an infinitesimal and radial displacement field $\xi(t,r)$ of a fluid element in the laboratory frame, and its coordinate-time
derivative $\dot\xi\equiv d\xi/d t$ gives the four-velocity of the fluid element, 
\begin{align}\label{uapprox}
U^\mu=(\gamma,\gamma\,\dot\xi,0,0), \quad \gamma=dt/d\tau.
\end{align}
The velocity $\dot\xi$ is related to the Lorentz gamma factor by $\gamma=(1-\dot\xi^2)^{-1/2}$. One can define so-called Eulerian perturbations (see for example Ref.~\cite{wheeler})
\begin{align}
 &\delta {\mathcal O}(t,r)\equiv {\mathcal O}(t,r)-{\mathcal O}_\mathrm{eq}(r)\label{eu}
\end{align}
for physical quantities ${\mathcal O}=(\n, \p, E)$. Eq.~(\ref{eu}) indicates by which amount these quantities differ at a certain space-time event $(t,r)$ in a perturbed configuration w.r.t.~the same
space-time event in the equilibrium configuration. These perturbations $\delta {\mathcal O}(t,r)$ are expressed in the coordinates of the laboratory frame. The thermodynamic quantities,
for example $\n$ and $\p$, and the laws of thermodynamics, are defined in the frame comoving with the respective fluid element to achieve covariance.
The Eulerian perturbations, for example $\delta \n$ and $\delta \p$, are associated with a fixed point in the laboratory frame.
We further introduce the so-called Lagrangian perturbations (see for example Ref.~\cite{wheeler})
\begin{align}\label{lagPert}
&\Delta {\mathcal O}(t,r)\equiv {\mathcal O}(t,r+\xi)-{\mathcal O}_\mathrm{eq}(r)\approx\delta {\mathcal O}(t,r)+\xi\,{\mathcal O}_\mathrm{eq}',\quad {\mathcal O}_\mathrm{eq}'\equiv d{\mathcal O}_\mathrm{eq}(r)/dr
\end{align}
for physical quantities ${\mathcal O}=(\n, \p, E)$. Eq.~(\ref{lagPert}) indicates by which amount these quantities of a perturbed fluid element $(t,r)$ moving with four
velocity $U^\mu=(\gamma,\gamma\,\dot\xi,0,0)$ differ from their counterparts of the same fluid element $(t,r)$ in the equilibrium configuration at rest in the laboratory frame.
In the following we will always assume the displacement field $\xi$ to be small enough that we are allowed to perform a Taylor expansion in powers of $\xi$ and apply the linear approximation of Eq.~(\ref{lagPert}).
In addition we will assume non-relativistic velocities $\dot \xi \ll 1$, then approximate $\gamma=1+O(\dot\xi^2)\approx 1$ by consistently neglecting the contribution of $O(\dot\xi^2)$.
This approximation is justified only for non-relativistic velocities $\dot\xi=\omega\,\xi\ll 1$ for stationary modes $\xi \propto e^{i\omega t}$ of eigen-frequency $\omega$. 
This condition imposes a further $\omega$-dependent constraint on the size of $\xi$, the linear approximation breaks down at large frequencies $\omega$ if $\xi$ is not chosen small enough.
The approximation is, however, justified also for relativistic equations of state, as long as the smallness condition is satisfied.

In the linear approximation, we calculate the Lagrangian perturbations $\Delta \n,\ \Delta \p$ and $\Delta E$ in terms of $\xi$. First we obtain $\Delta \n$ by using the
particle-number conservation (\ref{enNumCons})
\begin{align}\label{partNumCons}
 \nabla_\mu (\n U^\mu)=0\ \ \Longrightarrow \ \ U^\mu \nabla_\mu \n=\nabla_\tau \n=-\n \nabla_\mu U^\mu.
\end{align}
Using Eq.~(\ref{uapprox}) and the time-independence of the equilibrium configuration, Eq.~(\ref{partNumCons}) gives  
\begin{align}\label{Deltan}
 \Delta \dot \n=-\n_\mathrm{eq} \nabla_r \dot\xi\ \ \Longrightarrow \ \ \Delta \n=-\n_\mathrm{eq} \nabla_r \xi,
\end{align}
to the first order in $\xi$ and with appropriate integration constant. The definition of the adiabatic index $\Gamma_1$ of Eq.~(\ref{adiaInd}) provides us with a relation between $\Delta \n$ and $\Delta \p$ which reads
\begin{align}
 \Gamma_1\equiv\frac{\partial\ln\p}{\partial \ln\n}=\frac\n\p\frac{\Delta \p}{\Delta \n}.\nonumber
\end{align}
This implies the identity
\begin{align}\label{Deltap}
 \Delta \p=-\Gamma_1 \p_\mathrm{eq}\nabla_r \xi
\end{align}
in linear approximation. The Eulerian perturbations $\delta \n$ and $\delta \p$ follow from Eqs.~(\ref{Deltan}) and (\ref{Deltap}) by the use of Eq.~(\ref{lagPert}). The Eulerian perturbation of the electric field
$\delta E$ can be obtained from the $r$-component of Maxwell's equation (\ref{maxwell})
\begin{align}
 \nabla_\mu F^{\mu r}=\nabla_t (-E)=-4\pi\,e\,\gamma\,\n \,\dot\xi.\nonumber
\end{align}
Replacing $E\rightarrow E(r)+\delta E(t,r)$, dropping the Lorenz factor $\gamma$ in the linear approximation and integrating in time, we obtain
\begin{align}\label{eqn:eulerElec}
 \delta E=4\pi\,e\,\n\,\xi.
\end{align}
From Eq.~(\ref{eqn:eulerElec}) we can obtain the Lagrangian perturbation of the electric field as
\begin{align}
 \Delta \vec E = \delta \vec E + (\vec \xi\cdot \vec\nabla)\vec E,\nonumber
\end{align}
and in particular for the non-vanishing radial component
\begin{align}\label{eqn:lagElec}
 \Delta E = \delta E + \xi E'=\left(4\pi e \n_p - \frac{2}{r}E\right)\xi,
\end{align}
where the proton-number density $\n_p=3/(4\pi\Delta^3)\theta(r_c-r)$ . In Eq.~(\ref{eqn:lagElec}) we observe that the relative perturbation $\Delta E/E$ is finite, while $\Delta E$ and $E$ vanish at $r=r_\mathrm{max}$, as requested.
In linear approximation, we express the Euler equation (\ref{eulEqn}) in terms of 
\begin{align}
 \xi(t,r),\ \delta \n(t,r),\ \delta \p(t,r)\ \mathrm{and}\ \delta E(t,r),\nonumber
\end{align}
and obtain the following equation 
\begin{align}
 \ddot\xi=\frac{\partial_r\,\delta\p -e\,(\n_\mathrm{eq}\delta E + E_\mathrm{eq} \delta\n)}{\rh_\mathrm{eq}+\p_\mathrm{eq}}.\label{pEqn1}
\end{align}
Using Eqs.~(\ref{lagPert}) and (\ref{partNumCons}-\ref{eqn:lagElec}), Eq.~(\ref{pEqn1}) induces the second-order partial differential equation
\begin{align}\label{pertEqn}
 \ddot\xi=\frac{\partial_r(\Gamma_1 \p_\mathrm{eq}\nabla_r \xi+\xi\partial_r \p_\mathrm{eq})
   -4\pi e^2 \n_\mathrm{eq}^2\xi +e E_\mathrm{eq} (\n_\mathrm{eq}\nabla_r \xi+\xi\partial_r \n_\mathrm{eq})}{\rh_\mathrm{eq}+\p_\mathrm{eq}}
\end{align}
where we expressed all variations in terms of $\xi$ and the quantities $\n_\mathrm{eq}, \rh_\mathrm{eq}, \p_\mathrm{eq}$ and $E_\mathrm{eq}$ in a given equilibrium configuration
with fixed $r_c,Z$ and $r=r_\mathrm{max}$, as discussed in Sec.~\ref{thomFer}.
We are now in the position of finding stationary solutions $\xi(t,r)=\xi(r) e^{-i\omega t}$ to Eq.~(\ref{pertEqn}) by replacing $\ddot\xi\rightarrow -\omega^2\xi$.
We make use of Eqs. (\ref{fermiNumDens}-\ref{fermiEn}) and
(\ref{fullrelNumDens}) to express $\n_\mathrm{eq}$, $\rh_\mathrm{eq}$, and $\p_\mathrm{eq}$ in terms of the Thomas-Fermi function $\chi$. The equilibrium
electric field can be expressed as
\begin{align}
E_\mathrm{eq}=\frac{\chi}{r^2}-\frac{\chi'}{r},
\end{align}
which follows from (\ref{relThomFerAnsatz}). Then the perturbation Eq. (\ref{pertEqn}) takes the form
\begin{align}\label{eqn:Sturm-L}
 \left(\frac{A^{5/2}}{r^2 B}\xi'\right)'
 +\left(2 r^2\left(\frac{A^{5/2}}{r^5 B}\right)'
 +3 r A^{3/2}\left(\frac{\left(\chi/r\right)'}{r^2}\right)'
 -\frac{4\alpha A^3}{\pi r^4}\right)\xi=-\omega^2\frac{3 A^{3/2}B}{r^2}\xi
\end{align}
where we have used the abbreviations $A=2\,m_e r\,\chi + \chi^2$ and $B=m_e r + \chi$. The ultra-relativistic perturbation equation follows directly in the limit ($m_e\rightarrow 0$)
and reads
\begin{align}\label{eqn:Sturm-Lultra}
 \left(\frac{\chi^4}{r^2}\xi'\right)'
 +\left(2 r^2\left(\frac{\chi^4}{r^5}\right)'
 +3 r \chi^3\left(\frac{\left(\chi/r\right)'}{r^2}\right)'
 -\frac{4\alpha \chi^6}{\pi r^4}\right)\xi=-\omega^2\frac{3 \chi^4}{r^2}\xi.
\end{align}
Eqs. (\ref{eqn:Sturm-L}) and (\ref{eqn:Sturm-Lultra}) are Sturm-Liouville equations, so their normalized eigenfunctions for appropriate boundary conditions each form an orthonormal
basis with respect to the inner product that is based on the weight functions
$W_\mathrm{full}(r)=3 A^{3/2}B/r^2$ and $W_{\mathrm{ultra}}(r)=3\chi^4/r^2$, respectively,
\begin{align}\label{innerProd}
 \int_0^{r_\mathrm{max}} \xi_s(r)\xi_t(r) W(r)\mathrm{d}r=\delta_{s t}.
\end{align}
By varying $\omega$ to match boundary conditions, we will have to solve this boundary-value problem so as to obtain the eigen-frequencies $\omega_s$ corresponding to the eigen-modes $\xi_s$ of perturbation.
These boundary conditions are the following. First, the Lagrangian perturbation of pressure $\Delta \p\propto\nabla_r \xi=r^{-2}\partial_r (r^2 \xi)$ 
should vanish at the boundary ($r=r_\mathrm{max}$) of the configuration (see for example Ref.~\cite{wheeler}). This leads to the boundary condition
\begin{align}\label{pertOuterBound}
 \xi'(r_\mathrm{max})+\frac{2}{r_\mathrm{max}}\xi(r_\mathrm{max})=0.
\end{align}
Second, at the origin we have to demand that $\xi/r$
be finite or zero as $r\rightarrow 0$ in order for the perturbations to be finite (see for example Ref.~\cite{wheeler}). While we now have everything at hand that is needed to obtain the spectrum,
it turns out that numerical algorithms are not reliable in the inner region where $n=const.$ or $\chi\propto r$, respectively (see Figure \ref{fig:equilibriumPlot}). Under the very same condition,
however, we can easily solve the system analytically, as we will show in the next section. Then the analytical solution can be used to obtain boundary values for the numerical integration, taking place
from the region where densities start to vary, up to the radius of compression.
We make a consistency check that the condition $\xi_s(r) \omega_s \ll 1$ is fully satisfied in the entire region $[0, r_\mathrm{max}]$ of the system, indicating the validity of the linear approximation.
It is important to note that the eigen-frequencies $\omega_s$ we obtain by solving this boundary-value problem are completely independent of
$\xi(r_\mathrm{max})$-values chosen, provided the linear approximation is valid. However, the amplitudes of the Lagrangian perturbations $\Delta \n$, $\Delta \p$ and $\Delta E$ depend on
the $\xi(r_\mathrm{max})$-values chosen. We will discuss this in the concluding section.

\begin{figure}
 \centering
 \includegraphics[scale=0.7]{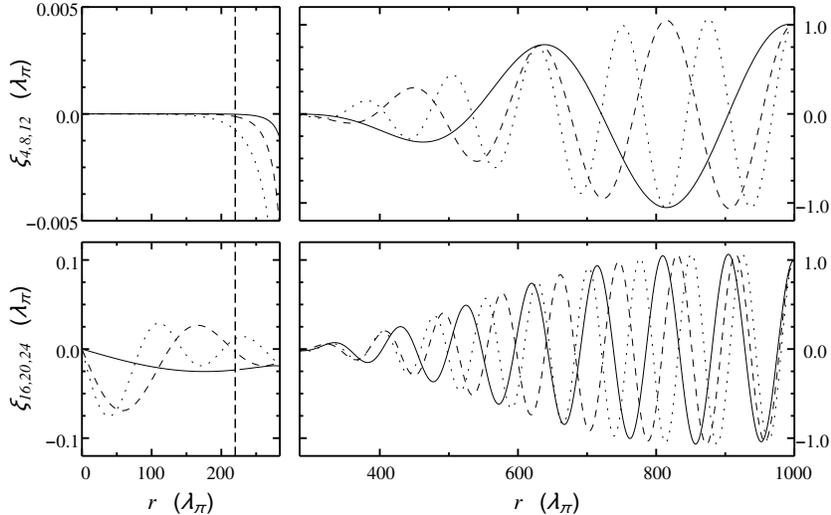}
 \caption{For $Z=10^6$, $\Delta=2.85$, $r_\mathrm{nuc}=285$, and $r_\mathrm{max}=1000$ the upper panels shows the displacement modes for $s=4,\,8,\,12$, and the lower panels for $s=16,\,20,\,24$
 (solid, dashed and dotted, respectively). The left panels cover the region
 inside the nucleus, while the right panels cover the region outside. The vertical dashed line indicates the point where analytical and numerical solutions are glued.}
 \label{fig:modesPlot}
\end{figure}

Before ending this section, it is worthwhile to mention that the pressure $\p$ does not vanish at the boundary $r_\mathrm{max}$, although its Lagrangian perturbation $\Delta \p$
vanishes. If there was no pressure $\p$ to compress the system, due to the screening effect of the Coulomb potential of protons and electrons, the system would extent to infinity,
so that numerically solving this boundary-value problem turns out to be impossible, as the spectrum becomes continuous. This is in contrast to the case of self-gravitating systems that have
finite boundary without compression due to the anti-screening effect of the gravitational potential of matter. 
In fact, the screening effect of electromagnetic interaction and the anti-screening effect of gravitational interaction are essentially different for electric pulsation and gravitational pulsation,
the former is stable and the latter can lead to unstable configurations \cite{chandra,wheeler}.

\section{\bf Obtaining the spectrum with a hybrid method}\label{exSol}

In this section we discuss the peculiar situation that arises if the equilibrium configurations of electrons have identical distributions to the positively charged nuclear core, namely 
a flat profile at nuclear density with $n=n_p=const.$ and $\chi\propto r$. In the ultra-relativistic framework this situation can be attained either exactly by a very high external pressure or approximately by the
strong electric force due to large proton-number $Z\gg 1$ and number-density at nuclear scale. The latter case is particularly interesting for the study of neutron star cores 
(see for example Refs.~\cite{rotondo2,rotondo3,rotondo,rueda}).  The reason why we are interested in this situation is that the exact solution of eigen-frequencies $\omega_s$ and eigen-functions $\xi_s$ to
this boundary-value problem for perturbations can be analytically obtained, so that we can discuss the influence of the plasma frequency on the resulting spectrum.
 \begin{figure}
 \centering
 \includegraphics[scale=0.7]{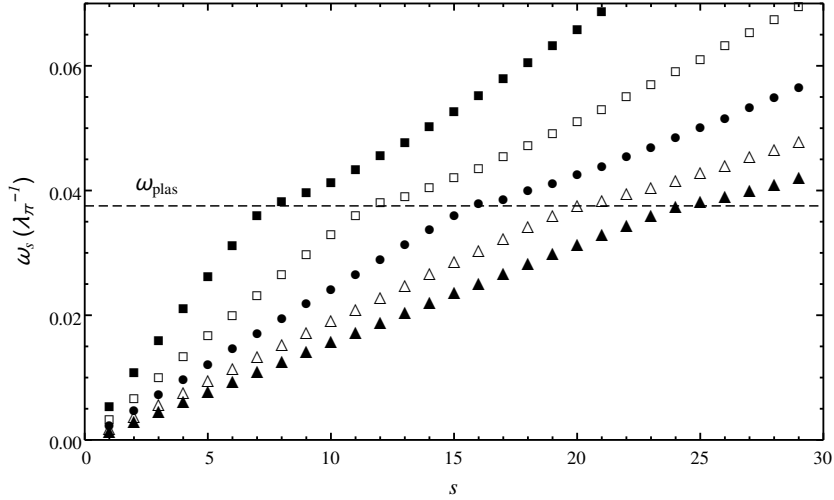}
 \caption{Spectrum of displacement modes for $Z=10^6$, $\Delta=2.85$, $r_\mathrm{nuc}=285$ and $r_\mathrm{max}=600$ (filled squares), $r_\mathrm{max}=800$ (empty squares), $r_\mathrm{max}=1000$ (filled circles),
 $r_\mathrm{max}=1200$ (empty triangles) and $r_\mathrm{max}=1400$ (filled triangles).}
 \label{fig:spectrumPlot}
\end{figure}
In these peculiar equilibrium configurations, the ultra-relativistic treatment is clearly justified because the electron number-density is near the nuclear density so that the electron Fermi-momentum is much larger
than the electron mass. Taking the ultra-relativistic limit of Eq.~(\ref{fullrelFermi}) and replacing the Heaviside function by unity, because the domain of interest is restricted to the interior of the
nucleus, we arrive at the equation
\begin{align}
 \frac{1}{3 r}\frac{d^2\chi}{d r^2}=-\frac{\alpha}{\Delta^3}+\frac{4\alpha}{9\pi}\frac{\chi^3}{r^3},
\end{align}
with boundary conditions (\ref{equilBoundCon}), where the radius of compression is set to the nuclear radius, $r_\mathrm{max}=\Delta\,Z^{1/3}$. The solution for the equilibrium configuration is simple and reads
\begin{align}\label{exactEquilSol}
 \n=\frac{3}{4\pi\Delta^3},\ \ \ \ \rh=\frac{9}{16\Delta^4}\left(\frac{3}{2\pi}\right)^{2/3},\ \ \ \ \p=\frac{3}{16\Delta^4}\left(\frac{3}{2\pi}\right)^{2/3},\ \ \ \ E=0,
\end{align}
i.e. the electrons are homogeneously distributed as are the protons, local neutrality follows, and electric charges cancel completely. Inserting (\ref{exactEquilSol}) into (\ref{pertEqn}) we find the second order differential equation
\begin{align}
 \xi''(r)+\frac{2}{r}\xi'(r)+\left[3\omega^2-\left(\frac{12}{\pi}\right)^{1/3}\frac{3\alpha}{\Delta^2}-\frac{2}{r^2}\right]\xi(r)=0\nonumber
\end{align}
for the perturbation of the electron fluid.
Invoking the first boundary condition at the origin $\xi(r\rightarrow 0)\rightarrow 0$ we see that the solutions must be of the form
\begin{align}
 \xi(r)={\mathcal C}\,j_{1}\left(
  \kappa\,r\right); \quad \kappa=\left[3\omega^2-\left(\frac{12}{\pi}\right)^{1/3}\frac{3\alpha}{\Delta^2}\right]^{1/2}\label{sol1}
\end{align}
where ${\mathcal C}$ is a constant and $j_{l}(z),\, l=0,1,2,\cdot\cdot\cdot$ denotes the spherical Bessel function of the first kind.
If we are interested in the case of compression up to the nuclear radius we have to fulfill the boundary condition (\ref{pertOuterBound}) and we
demand that $\kappa=\kappa_s = \pi s/r_\mathrm{max}$, leading to the eigen-frequencies 
\begin{align}
\omega_s=\left[\left(\frac{12}{\pi}\right)^{1/3}\!\!\frac{\alpha}{\Delta^2}+\frac{\pi^2 s^2}{3\,r_\mathrm{max}^2}\right]^{1/2};\quad  s=1,2,\cdot\cdot\cdot,\label{sol2}
\end{align}
and eigen-function $\xi_s(r)={\mathcal C}\,j_1(\kappa_s r)$ for the perturbation modes. The constant ${\mathcal C}$ is determined by the boundary value $\xi_s(r_\mathrm{max})$ and the trivial solution $s=0$ is excluded.
Thus we have obtained an analytical expression for electronic perturbation (pulsation) modes for the case that all electrons are compressed to the radius of the nuclear core. We can observe that the
spectrum is compsed of two parts: the first part containing the electromagnetic coupling constant $\alpha$ is an ultra-relativistic plasma frequency that would be present even if we were to consider pressure-less dust,
and the second part containing the wave number $\pi\,s/r_\mathrm{max}$ which is due to wave propagation in the medium.

As we have indicated in the last section, we can make use of the solution (\ref{sol1}) also in configurations that are not compressed up to the nuclear radius, because also these configurations are characterized by
a large volume in which electron and proton densities basically coincide. Because numerical methods do not work reliably in this regime, we obtain boundary values for the numerical integration from (\ref{sol1}) at a point just below
the surface of the nucleus, before the electron number density starts to change significantly, in our example at $r=220\,\lambda_\pi$. Then we glue the two solutions, the results can be seen in Figure \ref{fig:modesPlot}.

\begin{figure}
 \centering
 \includegraphics[scale=0.7]{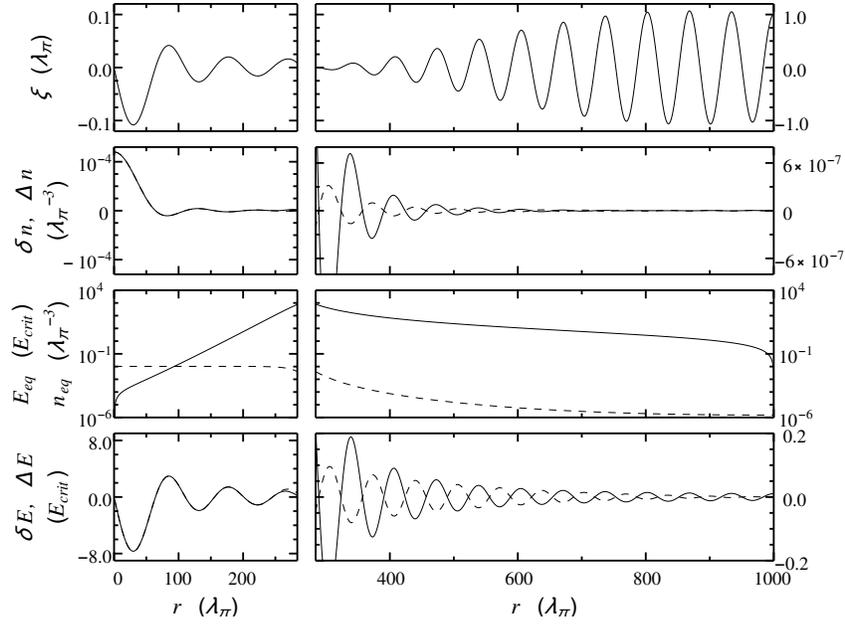}
 \caption{For $Z=10^6$, $\Delta=2.85$, $r_\mathrm{nuc}=285$, $r_\mathrm{max}=1000$ and $s=28$ we show from top to bottom displacement $\xi$, Eulerian (solid) and Lagrangian (dashed) number density perturbations
 $\delta n$ and $\Delta n$,
 equilibrium electric field (solid) and number density (dashed) $E_\mathrm{eq}$ and $n_\mathrm{eq}$, and Eulerian (solid) and Lagrangian (dashed) electric field perturbations
 $\delta E$ and $\Delta E$.}
 \label{fig:multiPanel}
\end{figure}

Note that in this case (in contrast to the full compression), the frequency $\omega$ can take values smaller than the plasma frequency $\omega_\mathrm{plas}=(12/\pi)^{1/3}\alpha/\Delta^2$, i.e. $\kappa$ can turn imaginary.
This leads to the displacement mode dropping off exponentially for small $s$. In Figure \ref{fig:modesPlot} we can observe this effect: because $\omega_{15}<\omega_\mathrm{plas}<\omega_{16}$, the modes in the upper left panel
are dropping off exponentially, while the modes in the lower left panel oscillate within the nucleus. The spectrum in Figure \ref{fig:spectrumPlot} further illustrates this effect: below $\omega_\mathrm{plas}$ the spectrum
has a steeper slope, because the volume of the nucleus is not available for propagation of the wave. Above $\omega_\mathrm{plas}$ the spectrum becomes less steep, because now also the volume of the nucleus
contributes to wave propagation. In Figure \ref{fig:multiPanel} we illustrate the Eulerian and Lagrangian perturbations of number density and electric field that arise for the perturbation mode corresponding to frequency $\omega_{28}$.

\section{\bf Conclusions and remarks.}\label{conClu}

Based on the approach of linear approximation, we quantitatively study the electric pulsation of the Thomas-Fermi equilibrium configurations around giant nuclear
cores with fixed proton number-density $n_p$ and compression radius $r_\mathrm{max}$. The eigen-frequencies $\omega_s$ and eigen-functions $\xi_s(r)$ of these electric perturbations are completely
determined, except the absolute amplitude of $\xi_s(r)$, corresponding to the arbitrary factor ${\mathcal C}$ in Eq.~(\ref{sol1}). The absolute amplitude of $\xi_s(r)$ has in general to be
determined by the space-time dependent strengths of external sources which initially trigger or continuously stimulate electric perturbations. In this case, the boundary-value problem turns out
to be inhomogeneous in Eq.~(\ref{pertEqn}) and/or boundary conditions (\ref{pertOuterBound}). This is the subject of study in our upcoming work \cite{hrx2014}, where we will use the spectrum
obtained here to solve the inhomogenous problem by means of the spectral method. In the present article, we have completely
determined the eigen-frequencies $\omega_s$ by choosing 
$\xi(r_\mathrm{max})$-values in such a way that the linear approximation is valid for 
$\omega_s\xi_s(r)\ll 1$. The amplitudes of eigen-functions $\xi_s(r)$,  as well as the corresponding Lagrangian perturbations of thermodynamical quantities
$\Delta \n$, $\Delta \rh$, $\Delta \p$ and electric field $\Delta E$ are determined up to a constant factor, like ${\mathcal C}$ in Eq.~(\ref{sol1}).  

\begin{figure}
 \centering
 \includegraphics[scale=0.7]{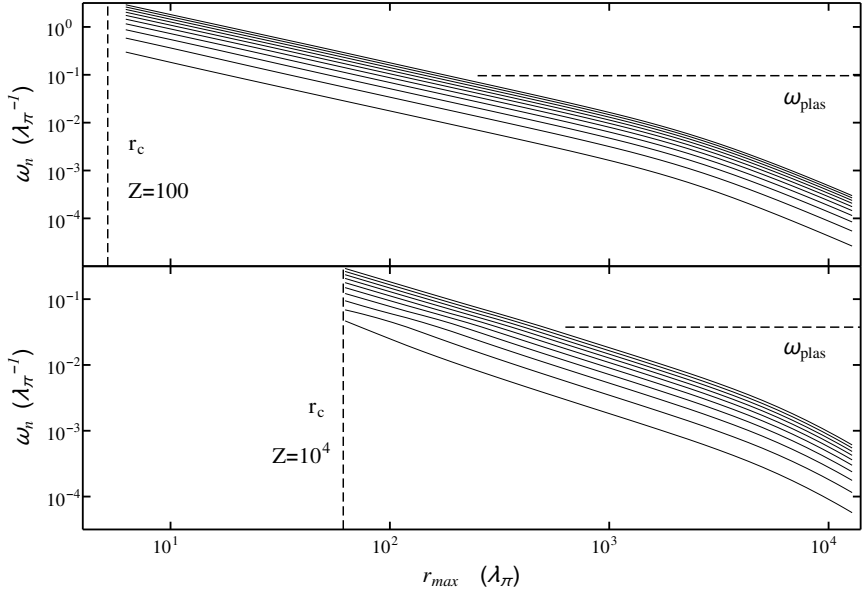}
 \caption{The eigen-frequencies $\omega_{1-10}$ for $Z=100$ and $A=2 Z$ (upper panel), and for $Z=10^4$ and saturated $\beta$-equilibrium (lower panel) versus the radius of compression.
 In the former case we can only observe the transition from non- to ultra-relativistic equation of state, in the latter case the influence of the nuclear plasma frequency becomes
 visible: for frequencies above $\omega_\mathrm{plas}$ the volume of the nucleus becomes available for wave propagation, rendering the slope less steep.}
 \label{fig:10to2and4SpectPlot}
\end{figure}

Naturally the question arises whether the eigen-frequencies $\omega_s$, in particular the lowest-lying one $\omega_1$, can be experimentally tested, since these eigen-frequencies $\omega_s$
are characteristic frequencies (time scales) of the Thomas-Fermi system responding to suitable external actions.
To give an impression how the spectrum of our model behaves over a large range of compression radii $r_\mathrm{max}$ we have integrated the full relativistic
Eqs. (\ref{fullrelFermi}) and (\ref{eqn:Sturm-L}) for $Z=100$ and
 $Z=10^4$ with suiting approximate values for $\beta$-equilibrium (see Table \ref{tab:freq}) and show the results in Figure \ref{fig:10to2and4SpectPlot}.
It becomes clear from this figure that, while the transition from the non- to the ultra-relativistic equation of state is visible in the upper panel,
which corresponds to common nuclei that could be probed in laboratory experiments, the constribution
of the plasma frequency is negligible. In the lower panel, however, which corresponds to systems that might be expected in the lower crust of neutron stars, there is a visible effect of
the plasma frequency on the spectrum at compression radii that are to be expected in such systems.
In order to simplify the boundary-value problem and numerical calculation, we adopt
a spherically symmetric model and find the eigen-frequencies and eigen-functions for spherical electric pulsations in the radial direction. Such spherical electric pulsations
(the monopole vibrations) do not emit electromagnetic radiation (spin-one) due to the conservation of angular momentum. However, if structures in the lower crust such as the ones present
in a \textit{pasta equation of state}, or neutron star cores, do not possess the exact spherical
symmetry, the dipole-component of electric pulsations appears, and this possibly leads to electromagnetic radiation with discrete frequencies.
Such effects could be introduced to our spherically symmetric system
by dropping the symmetry assumption for the perturbation, and expanding in the appropriate basis, e.g. spherical harmonics instead. In the context of the \textit{pasta state} also
other geometries such as axial and planar symmetries could be taken into consideration for the equilibrium state, with most of the calculation remaining unchanged. This procedure would allow to predict emission of electromagnetic radiation
at frequencies to be determined by the details of the calculation (geometry, pressure), stimulated by dynamic events in the neutron star such as slowdown/spinup glitches or even more powerful events such as neutron star mergers.
Furthermore, the effects we observe should influcence the dynamic compressibility of nuclear matter in the \textit{pasta state}, but it is not clear when the tools will be available to probe such effects. If gravitational
wave astronomy can be advanced to a precision that allows to probe tidal effects in neutron star mergers, it will allow to constrain crust equations of state.

The analytical pulsation frequencies obtained for the peculiar equilibrium configuration in Sec.~\ref{exSol} could find their application in the context of neutral nuclear matter at/over
nuclear density, for instance neutron star cores.
The electrons and protons approximately have the same density profiles and 
occupy the same volume, except the very thin layer on the surface of neutron star cores \cite{rotondo, rotondo2,rotondo3,rueda}, so the numerical results obtained for $Z=10^6$ can give insight into the pulsation behaviour
near this surface.
Moreover, if the amplitude of a perturbation
is so large that the perturbation of the electric field is over the critical value $E_c=m_e^2c^3/(e^2\hbar)$, electron-positron pairs are produced, leading to some observational consequences
(see Ref.~\cite{hrx2012, rx2013}).

\begin{table}
\begin{tabular*}{\textwidth}{lccc}
 \hline 
$Z$ &\ \ \ \ \ \ \ \ \ \ $10^2$\ \ \ \ \ \ \ \ \ \ &\ \ \ \ \ \ \ \ \ \ $10^4$\ \ \ \ \ \ \ \ \ \ \ \ &$10^6$\ \\ 
$A$ &$2\times10^2$\ &\ \ $2.2\times 10^6$\ \ &\ \ $2.2\times 10^8$\ \ \\ \hline
$r_\mathrm{max}\ \ \ \ \ (\lambda_\pi)$\ \ \  & 8000 & 125 & 285\\
$\langle n_e \rangle\ \ \ \ \ \,(\lambda_\pi{}^{-3})$\ \ \  & \ \ $5\times 10^{-11}$ & $10^{-3}$ & $10^{-2}$\\
$\omega_1$\ \ \ \ \ \ \ \,(MeV)\ \ \ &\ \  $9\times 10^{-3}$\ \  & 2.6\ \  & 5.3\\
$\omega_\mathrm{plas}$\ \ \ \ \,(MeV)\ \ \  & 13\ \  & 5.3\ \  & 5.3\\
$\omega_\mathrm{breath}$\ \ (MeV)\ \ \  & 39\ \  & 11\ \  & 5.2\\ \hline
\end{tabular*}
\caption{Comparison for suitable radius of compression $r_\mathrm{max}$ of average electron density $\langle n_e \rangle$, corresponding fundamental mode $\omega_1$,
plasma frequency $\omega_\mathrm{plas}$ and nuclear breathing frequency $\omega_\mathrm{breath}$. Based on the assumptions of nuclear density,
bulk nuclear incompressiblity of $K_0=220\ \mathrm{MeV}$, $Z=A/2$ for $Z=10^2$ and saturated $\beta$-equilibrium in the other cases. The proton density $n_p$ in saturated $\beta$-equilibrium
equals $\langle n_e \rangle$ in the third column.}
\label{tab:freq}
\end{table}

To end this article, we briefly discuss the possibilities of external perturbations from nuclear cores, which probably cause the phenomenon of electronic pulsations discussed in this article.
As an example, in Ref.~\cite{wang}, the nuclear breathing
modes of nuclear collective motion at the time scale of the nuclear force were discussed. We attempt to model the perturbation of nuclear cores and solve the inhomogeneous boundary-value problem
to have further understanding of electric pulsations around nuclear matter \cite{hrx2014}. For this pupose we make a rough estimate and consider only the bulk nuclear
incompressibility $K_0\approx 220\,\mathrm{MeV}$ as it dominates for large $A$. The nuclear breathing frequency is calculated according to \cite{wang} as
\begin{align}
 \hbar\,\omega_\mathrm{breath}=\sqrt{\frac{\hbar^2 K_0}{m_{_N} \langle r^2 \rangle}}=\sqrt{\frac{3 \hbar^2 K_0}{5 m_{_N} r_\mathrm{c}^2}}
\end{align}
where $m_{_N}$ is the neutron mass, $r_c$ the radius of the nuclear core, and we have assumed homogenous distribution of nuclear matter in the core.
In Table \ref{tab:freq} we compare the nuclear breathing modes to the plasma frequencies. While the former depend on the size of the nucleus, the latter depend only on proton density, and we observe
that they are of comparable size for $Z\approx 10^6$. As we have outlined in \cite{zeldoProc} a driving force proportional to the radius, corresponding to a homogeneous nuclear breathing mode,
can easily be modelled using the spectral method and the inner product (\ref{innerProd}), effectively calculating time-dependent coefficients for the pulsation modes derived in this article.

It is worthwhile to mention that in gravitational collapses of macroscopic cores at/over nuclear density, strong dynamical variations of collapsing cores can be induced at or over the rate of
nuclear interactions, which cannot be treated as linear perturbations, and these variation are no longer stationary. Non-perturbative calculations showed that this results in strong variations of the electric field,
leading to the formation of electron-positron pairs and photons at high energy- and number-densities \cite{hrx2012,rx2013}. Further quantitative studies on this issue are being conducted. 

\section*{Acknowledgements}
H. Ludwig is supported by the Erasmus Mundus Joint Doctorate Program under Grant
Number 2012-1710 from the EACEA of the European Commission.

\end{document}